\documentclass[prb,twocolumn,showpacs,preprintnumbers,amsmath,amssymb]{revtex4-1}

\usepackage{graphicx,rotating}
\usepackage{dcolumn}
\usepackage{bm}
\usepackage{epsfig}
\usepackage{longtable}
\usepackage{color}

\begin{document}

\title{Ferrimagnetism in 2D networks of porphyrin-X and -XO (X=Sc,...,Zn)
with acetylene bridges.}

\author{Ma\l{}gorzata~Wierzbowska\email{wierzbowska@ifpan.edu.pl}
and Andrzej~L.~Sobolewski}

\affiliation{Institut of Physics, Polish Academy of Science (PAS),
Al. Lotnik\'ow 32/46, 02-668 Warszawa, Poland}

\date{\today}

\begin{abstract}
Magnetism in 2D networks of the acetylene-bridged transition metal porphyrins 
M(P)-2(C-C)-2 (denoted P-TM), and oxo-TM-porphyrins 
OM(P)-2(C-C)-2 (denoted P-TMO), 
is studied with the density functional theory (DFT)   
and the self-interaction corrected pseudopotential scheme (pSIC).  
Addition of oxygen lowers magnetism of P-TMO with respect to the corresponding P-TM 
for most of the first-half $3d$-row TMs.   
In contrast, binding O with the second-half $3d$-row TMs or Sc increases 
the magnetic moments.
Ferrimagnetism is found for the porphyrin networks with the TMs from V to Co
and also for these cases with oxygen.  
This is a long-range effect of the delocalized spin-polarization, 
extended even to the acetylene bridges. 
\end{abstract}


\maketitle

\section{Introduction} 

Porphyrins are popular molecules in living organisms. Their ability to 
bind the transition metals is utilized in biological processes significant 
for the human-blood functions - employing Fe in heme. 
It is also responsible for the photosynthesis in plants -
building a complex with Mg, namely chlorofile.\cite{biol}
The oxo-metal-porphyrins, such as oxotitanium porphyrin (P-TiO),  
can be used for solar-activated water splitting - 
what has been theoretically proposed to occur due to low-lying ligand-to-metal
intramolecular charge-transfer states\cite{pccp1} and
recently experimentally confirmed.\cite{pccp2}

Porphyrins or phthalocyanines can be assembled at metals\cite{jcp,grow2014} 
and other surfaces\cite{grafen} forming 2D networks.\cite{metal-org} 
In turn, the surfaces or additional ligands change 
the charge state of the transition metal
bound to porphyrins with respect to the bare 2D network. This is because 
different valence states of a metal, metal-oxygen moiety, or metal-Li complex 
lay closely in the energy.\cite{broclawik,porf-Cr-Co,mixed,metal-Li}   
Not only the single layers of the metal-organic networks exist, 
also the double- and triple-decker layers of metal-phthalocyanines have been grown,
on both the ferro- and antiferromagnetic substrates.\cite{decker}

For catalytic reactions with the intramolecular charge transfer,
such as the water splitting, the intermolecular connections must be 
very weakly conducting (non-covalent);
for example the hydrogen bridges (O-H...O).
For the spintronic applications, on the other hand, 
these connections should be conducting. For this reason, we
have chosen the acethylene bridges.
Various choices of the intermolecular-bridging type were examined experimentally
for the energy transfer rates,\cite{eET1,eET2} and a role of the  frontier
molecular-orbitals was theoretically studied for the donor-acceptor dimers.\cite{frontier} 

A new branch of material science grows on top of potential utilization  
of the 2D organic layers\cite{new} in the electronic devices;
as well as the magnetic nanoparticles deposited on various substrates.\cite{nanomagn} 
Selective incorporation of metal atoms into organic templates 
enables wide functionalization of the layers,\cite{function}
hence many applications are plausible.  
The 4-fold symmetry of porphyrins and phthalocyanines with the transition metals  
makes their networks similar to the Heusler compounds, 
which are very strong magnets.\cite{heusler}
There are review articles on the spintronic topics in the 2D networks.\cite{mrs-1,mrs-2} 
The photovoltaic thin films based on the metal-phthalocyanine 2D networks - organized in 
the AA-stacking order - conduct the current very well in the columnar direction.\cite{photovolt} 
Similar effect should be observed in metal-porphyrins.
For technological reasons, the conductive properties of the single metal-porphyrin molecules
between the electrodes,\cite{Zn-cond} and the 1D metal-porphyrin chains\cite{1D-porf} 
were also investigated. Recently, high-mobility field-effect transistors
have been constructed by building the ABAB-type stacking of 
the phthalocyanine-VO and -TiO layers.\cite{transistor}

In this work, we focus on the covalent 2D networks of porphyrins with 
all $3d$-row transition metals and TMO, bridged with the acetylene moiety.  
The model structures of our systems are presented in Fig.~1. 
We want to obtain a material with strong and long-range magnetic-exchange 
interactions. Thus the covalent intermolecular connection was chosen,  
since it has been found that the covalent inorganic materials - as hosts
for the transition metal impurities - are strongly magnetically coupled for 
long distances.\cite{siretc,maggraf} 

\begin{figure}[ht]
\centerline{
\includegraphics[scale=0.17,angle=0.0]{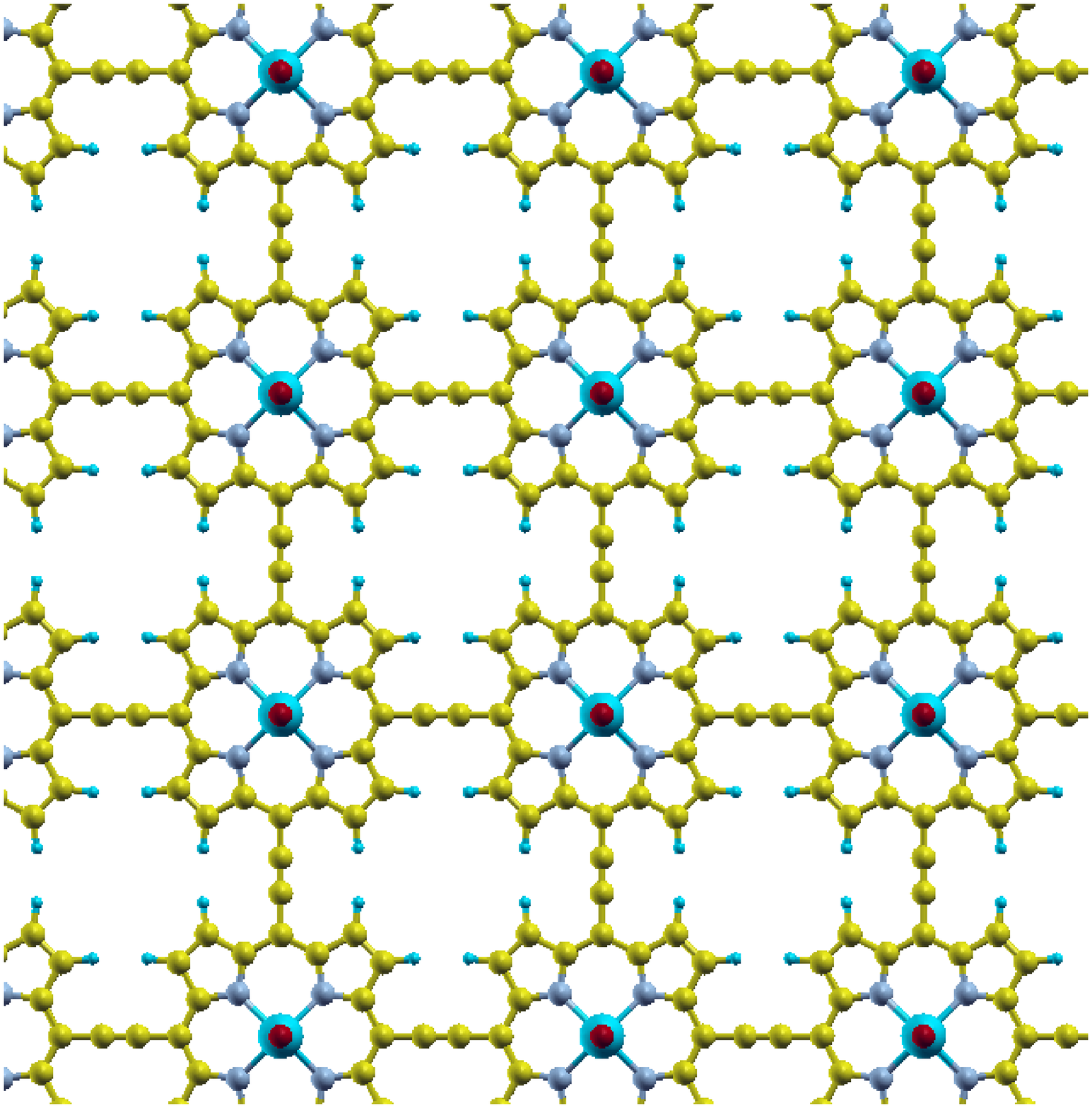}\hspace{2mm}
\includegraphics[scale=0.17,angle=0.0]{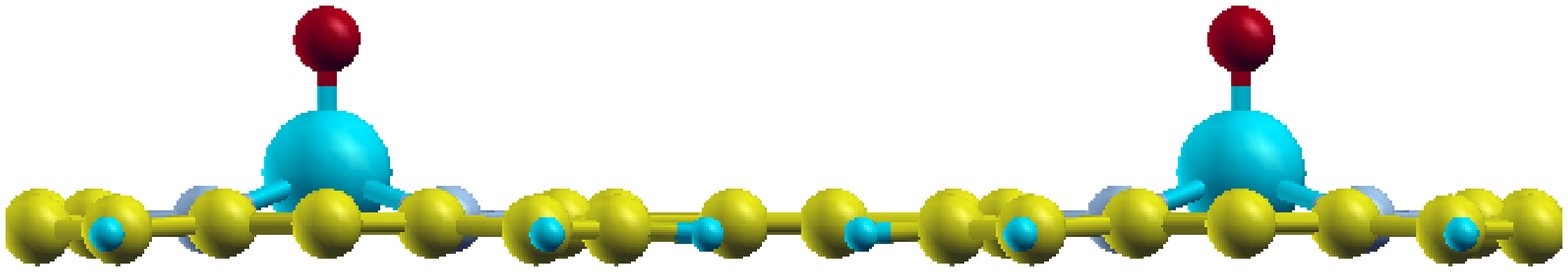}}
\caption{2D network of porphyrin-MnO with acetylene bridges; top view and side view.}
\label{mol}
\end{figure}

Interestingly, another molecular system - similar to that studied in this work, 
but assembled at Cu and containing Fe - was demonstrated to switch the easy
magnetization-direction after adding the oxygen.\cite{FeO-axis}
We report the ferrimagnetism, similar to that theoretically predicted
by other authors in porphyrins with Mn 
and connected by 4-bromophenyl to form 1D magnetic chains.\cite{ferrimag}
This effect can be mapped on the atomic scale by the measurement of magnetic
resonance spectra at gigahertz frequencies using X-ray magnetic circular
dichroism (XMCD).\cite{xmcd}

\section{Theoretical details} 

We performed the density-functional theory (DFT) calculations using  
the {\sc Quantum ESPRESSO} (QE) suite of codes,\cite{qe} which employs the 
plane-wave basis set and the pseudopotentials to describe the core electrons.
The exchange-correlation functional was chosen for the gradient corrected
Perdew-Burke-Ernzerhof (PBE) parametrization.\cite{pbe} 
The ultrasoft pseudopotentials
were used with the energy cutoffs 35 Ry and 400 Ry for the plane-waves and
the density, respectively. 
The Monkhorst-Pack uniform k-mesh in the Brillouin zone (BZ) has been set to
$10\times 10\times 1$, and the Fermi-surface energy broadening parameter 
0.02 Ry was chosen for a better convergence. 
The vacuum separation between the periodic slabs was 40 \AA. 
The pseudopotentials for the atoms from Sc to Co were modeled with 
the semi-core states (3s and 3p) included in the valence band,
in contrast, the Cu and Zn valence shells were constructed 
without the deeper states. 

We started with the geometry optimization of a single molecule, 
namely porphyrin-Ti,
using the B3LYP method which is equivalent to the DFT scheme of the BLYP-type 
with 20$\%$ of the exact exchange  (EXX).\cite{b3lyp-1,b3lyp-2,b3lyp-3}
This step was done with the quantum chemistry package TURBOMOLE,\cite{turbo}
which represents the wavefuntions in the gaussian basis set; we used 
the correlation-consistent valence double-zeta atomic basis set
with polarization functions for all atoms (cc-pVDZ).\cite{cc-pvdz}
Obtained geometry and the lattice vectors, derived from the molecular size,  
were inserted as an input for the geometry optimization of the periodic structures, 
calculated by means of the generalized gradient approximation (GGA) scheme 
(with the PBE parametrization) using the QE tool.  
The above procedure is accurate enough, since the central 'squares' of porphyrin 
and phthalocyanine derivatives do not relax much after binding a metal.\cite{teochem}

It is known, that the DFT approach underestimates the energy gaps and 
the energetic positions, with respect to the Fermi level, 
of the localized d- or f-shells.
This fact has consequences in the description of magnetization. 
In order to improve the treatment with the lack of the exact exchange, 
we applied the pseudopotential self-interaction correction (pSIC) method 
proposed by Filippetti and Spaldin,\cite{psic-1} 
and implemented by us in the QE package.\cite{psic-2}
The pSIC method is superior to the DFT+U approach\cite{dft+U} for two reasons: 
(i) the correction is parameter free, unlike the DFT+U parameters: 
the Coulomb U and the exchange J, 
(ii) the correction is applied to all atomic shells, 
not only d- or f-shell of the transition metals or the rare earths.
The pSIC-kernel includes the Hartree and the exchange-correlation potentials, 
$ V_{HXC}^{\sigma}$, 
calculated on the orbital-density $n_i^{\sigma}({\bf r})$, dependent on spin $\sigma$, 
and built using the atomic pseudopotential orbital, 
$\varphi_i({\bf r})$. 
The main equations are: 
\begin{eqnarray}
\hat{V}_{SIC}^{\sigma} & = & \sum_{i} \; \frac 
{| \varphi_i({\bf r}) V_{HXC}^{\sigma}[n_i^{\sigma}({\bf r})] \rangle \;
\langle V_{HXC}^{\sigma}[n_i^{\sigma}({\bf r})] \varphi_i({\bf r}) |}
{\langle \varphi_i({\bf r}) | V_{HXC}^{\sigma}[n_i^{\sigma}({\bf r})] | 
\varphi_i({\bf r}) \rangle }  \nonumber \\
n_i^{\sigma}({\bf r}) & = & p_i^{\sigma} \; | \varphi_i({\bf r}) |^2  \nonumber \\
p_i^{\sigma} & = & \sum_{n{\bf k}} \; f_{n{\bf k}}^{\sigma} \; 
\langle \psi_{n{\bf k}}^{\sigma} | \varphi_i \rangle
\langle \varphi_i | \psi_{n{\bf k}}^{\sigma} \rangle \nonumber
\end{eqnarray}
The occupation numbers $p_i^{\sigma}$ are obtained from the projection of 
the Kohn-Sham states
$\psi_{n{\bf k}}^{\sigma}$ onto the pseudopotential atomic-orbitals $\varphi_i({\bf r})$  
and $f_{n{\bf k}}^{\sigma}$ are the Fermi-Dirac occupations. 
Usefulness of this method has been demonstrated in a number
of various applications.\cite{psic-3,sire,hole} 
For this work, 
the most important are the relations between the energetic positions of 
the shells: $3d$-TM, $4s$-TM and $4p$-TM, $2p$-O and $2p$-N. 
Although, we included the SIC for all atomic shells, these corrections are 'on-site'
- the same like in the DFT+U approach and contrary to the Nagaoka model, which includes
the 'inter-site' strong correlations on the parametric grounds.\cite{nagaoka-1,nagaoka-2}  
Using the pSIC approach instead of the DFT+U scheme is especially justified for the
molecular crystals and 2D metal-organic frameworks, due to the fact they possess the flat
band structures.\cite{flat} In constrast, the ordinary strongly correlated systems such
as the metal oxides and diluted magnetic semiconductors possess only $d$-type or $f$-type flat
bands and the rest of projected atomic states have usual semiconducting bandwidths.  

\section{Results and discussion} 

\begin{figure*}[ht]
\includegraphics[scale=0.4,angle=0.0]{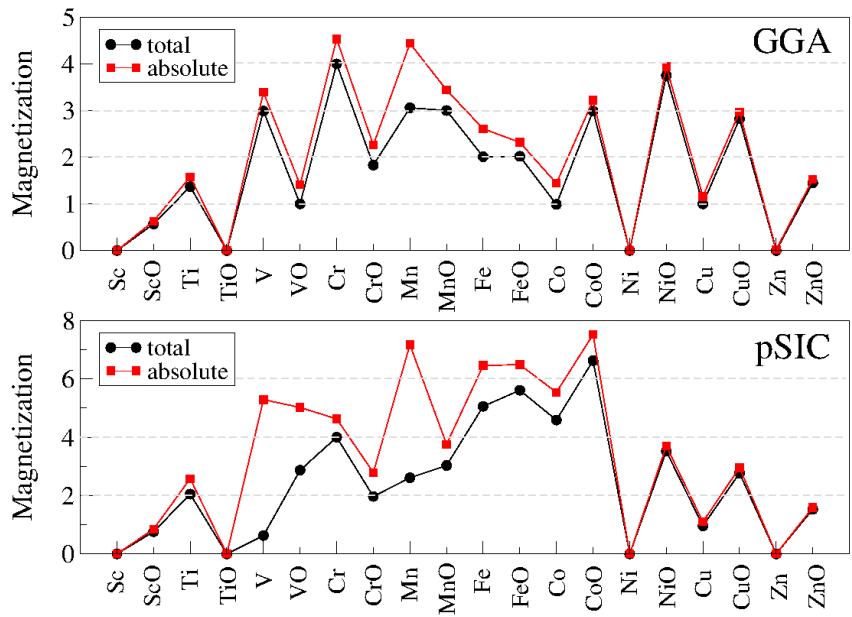}
\caption{Total and absolute magnetizations 
(in $\mu_B$) for metal-porphyrins and oxo-metal-porphyrins
with the acethylene bridges,
obtained with the GGA and pSIC approaches using the GGA-optimized geometry.}
\label{magnet}
\end{figure*}

In this study, we focus our attention on magnetism, 
which is a phenomenon sensitive to the system geometry. 
As mentioned in the theoretical details section,
the geometry was optimized for all studied cases. 
Positions of transition metal atoms above the plane of the nitrogens square 
and the oxygen-TM bond lengths are collected in the Table~S1 
in the supplemental information.\cite{supp}
Inspecting the atomic gradients varied during the geometry optimization
and the corresponding magnetizations of the systems, 
we noticed that a tiny change of the TM-positions
or the TM-O bond-lengths can influence magnetizations. 
This effect is similar to the phenomenon observed in the perovskites\cite{perovskity} 
and also for metal atoms at graphene and graphite.\cite{at-graf} 
In our cases, however,
the geometric effect does not alter main trends in the results reported below. \\ 

\subsection{Magnetizations}

In Fig.~2, the total and absolute magnetizations in the elementary cells, 
obtained within the GGA and the pSIC frameworks,
are presented for porphyrins with all TM and TMO additions.  
The total magnetization, defined via the up and down spin-densities,  
$n_{\uparrow}({\bf r})$ and $n_{\downarrow}({\bf r})$, as
\begin{equation}
M_{total} \; = \; \int \; n_{\uparrow}({\bf r}) \; d{\bf r} \; - 
                  \int \; n_{\downarrow}({\bf r}) \; d{\bf r},  \nonumber
\end{equation}
reflects the summed magnetization of the sample seen from a distance.   
In contrast, the absolute magnetization, defined as
\begin{equation}
M_{absolute} \; = \; \int \; |\; n_{\uparrow}({\bf r}) - 
n_{\downarrow}({\bf r}) \; | \; d{\bf r},
 \nonumber
\end{equation}
means a sum of the local magnetizations regardless their signum.
Large difference between the total and the absolute magnetizations gives  
an information on the space separation of the alterred magnetic moments,
i.e. the ferrimagnetic character of the sample. 

Ferrimagnetism is well pronounced for the porphyrins with metals from V to Co,
and it is the strongest in the Mn, Fe and FeO cases 
for both calculation methods, GGA and pSIC, and additionally in 
the V and VO cases obtained with the pSIC scheme.
The spin-density map for the porphyrin-Mn network is presented 
in Fig.~3 for the pSIC method.
The vanadium spin-density map is similar and included in Fig.~S2 
in the supplemental information.\cite{supp}
It is clear that the nitrogen atoms are spin-polarized in the TM vicinity. 

The effect of oxygen addition on magnetization is very strong. 
In the case of the TM atoms from the first half of the $3d$-row,  
the oxygen addition strongly reduces the magnetic moments; except the porphyrin-ScO network. 
In contrast, for the TM atoms from the second half of the $3d$-row, addition of oxygen  
causes an increase of the magnetic moments. 
This is an interesting result - 
similar to the results for co-doping of metal-phthalocyanines with Li\cite{metal-Li} -
and might be utilized in spintronic devices. \\

\begin{figure}[ht]
\includegraphics[scale=0.3,angle=0.0]{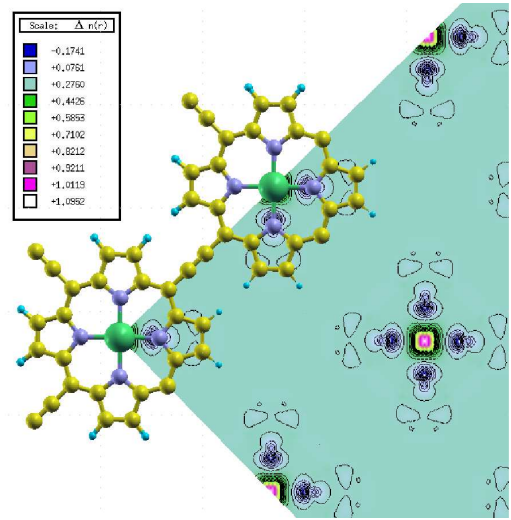}
\caption{Spin-density map of porphyrin-Mn, obtained with the pSIC approach.} 
\label{mapa}
\end{figure}

\begin{figure}[ht]
\includegraphics[scale=0.23,angle=0.0]{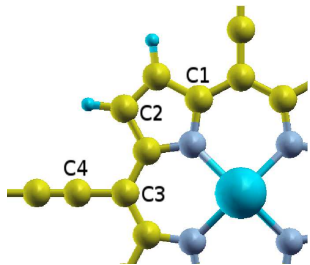}
\caption{The indices for carbon atoms used in the Table~1.}
\label{index}
\end{figure}

\begin{table*}
\caption{L\"owdin spin-polarizations (in $\mu_B$) for the 2D porphyrin 
networks with TMO (upper part) and TM (lower part); calculated with 
the pSIC scheme. Different carbon atoms are indexed as follows: 
C1 denotes carbons adjacent to N,
C2 carbons terminated with H, C3 carbons adjacent to acetylene, and 
C4 carbons of acetylene - they are presented in Fig.~4. 
The corresponding GGA numbers are given in parenthesis.}
\begin{tabular}{lcccccccccc}
\hline \\[-0.1cm]
shell   & $\;\;\;$  Sc $\;\;\;\;\;$  & $\;\;\;\;\;$ Ti 
$\;\;\;\;\;$ & $\;\;\;\;\;$ V $\;\;\;\;\;$ & 
        $\;\;\;\;\;$ Cr $\;\;\;\;\;$ &  $\;\;\;\;\;$ Mn 
$\;\;\;\;\;$ & $\;\;\;\;\;$ Fe $\;\;\;\;\;$  &  
        $\;\;\;\;\;$ Co $\;\;\;\;\;$ & $\;\;\;\;\;$ Ni 
$\;\;\;\;\;$ & $\;\;\;\;\;$ Cu $\;\;\;\;\;$ & 
        $\;\;\;\;\;$ Zn $\;\;\;\;\;$ \\[0.1cm]
\hline \\[-0.2cm]
O $sp$ & 0.720 & 0.0 & -1.570 & -0.215 & 0.043 &  1.670 &  2.265 & 1.546 &  1.636 &  1.491 \\
      & (0.577) & (0.0) & (-0.149) & (-0.062) & (0.577) & (0.728) & (1.392) & (1.403) & (1.435) & (1.389) \\[0.1cm]
TM $d$ & 0.003 & 0.0 &  0.940 &  2.036 & 2.991 &  2.612 &  3.026 & 1.562 &  0.506 &  0.001 \\
      & (-0.016) & (0.0) & (1.154) & (1.684) & (2.424) & (1.128) & (1.579) & (1.749) & (0.677) & (0.025) \\[0.1cm]
TM $sp$ $\;\;\;$ & 0.007 & 0.0 & 0.005$^1$ & 0.038 & 0.087 &  0.015 & 0.008  & -0.005 & -0.044 & -0.006 \\
       & (-0.001) & (0.0) & (0.020) & (0.021) & (0.024) & (0.006) & (0.013) & (-0.010) & (-0.043) & (-0.009) \\[0.1cm]
N $sp$ & 0.000 & 0.0 & -0.270 & -0.047 & -0.036 &  0.051 &  0.057 & 0.101 &  0.158 &  0.008 \\
      & (0.004) & (0.0) & (-0.019) & (-0.032) & (-0.030) & (-0.007) & (0.003) & (0.135) & (0.164) & (0.011) \\[0.1cm]
C1 $sp$ & 0.000 & 0.0 & -0.120 &  0.024 & 0.006 &  0.109 &  0.117 & -0.005 & -0.007 & -0.001 \\
C2 $sp$ & 0.000 & 0.0 & -0.057 &  0.007 & 0.001 &  0.056 &  0.047 & 0.007 &  0.007 &  0.001 \\
C3 $sp$ & 0.003 & 0.0 &  0.052 &  0.003 & -0.005 & -0.051 & -0.057 & -0.004 &  0.007 &  0.001  \\
C4 $sp$ & 0.001 & 0.0 &  0.006 &  0.003 & -0.001 & -0.006 & -0.003 & -0.003 &  0.001 &  0.001 \\[0.1cm]
\hline \\[-0.2cm]
TM $d$ & 0.0 &  1.570  &  1.630 &  3.820 &  4.158 &  3.940 &  3.046 & 0.0  &  0.420 &  0.0 \\
     & (0.0) & (0.989) & (2.510) & (3.656) & (3.543) & (2.092) & (-1.044) & (0.0) & (0.490) & (0.0) \\[0.1cm]
TM $sp$ & 0.0 & 0.014 & -0.971$^2$ & 0.140 & -0.235$^3$ & 0.074 &  0.046 &  0.0  & -0.027 &  0.0 \\
      & (0.0) & (0.199) & (0.197) & (0.163) & (0.146) & (0.013) & (-0.092) & (0.0) & (-0.018) & (0.0) \\[0.1cm]
N $sp$ & 0.0 & -0.063 & -0.290 & -0.074 & -0.470 & -0.109 &  0.062 &  0.0 &  0.136 &  0.0 \\
      & (0.0) & (-0.019) & (-0.057) & (-0.068) & (-0.044) & (-0.036) & (0.011) & (0.0) & (0.129) & (0.0) \\[0.1cm]
C1 $sp$ & 0.0 &  0.057 &  0.210 &  0.034 &  0.093 &  0.172 &  0.109 & 0.0  & -0.004 &  0.0 \\
C2 $sp$ & 0.0 &  0.027 & -0.019 &  0.010 &  0.009 &  0.050 &  0.079 & 0.0  &  0.005 &  0.0 \\
C3 $sp$ & 0.0 &  0.002 & -0.002 & -0.009 & -0.039 & -0.071 & -0.056 &  0.0 &  0.002 &  0.0 \\
C4 $sp$ & 0.0 &  0.003 & -0.095 & -0.004 & -0.004 & -0.004 & -0.007 &  0.0 &  0.000 &  0.0 \\
\hline \\[-0.2cm]
\multicolumn{4}{l}{$^1$  0.032(s), -0.27(p)} & 
\multicolumn{4}{l}{$^2$  0.006(s), -0.977(p)} & 
\multicolumn{3}{l}{$^3$  0.099(s), -0.334(p)} \\
\end{tabular}
\end{table*}

To get further insight into the electronic structure and magnetism, it is useful to
examine the L\"owdin population numbers. 
The differences between the L\"owdin population numbers for the
spin up and spin down are the spin polarizations of chosen
atomic orbitals. They are listed in Table~1, for all porphyrins with TM and TMO.
We present the numbers obtained mainly with the pSIC scheme. 
The corresponding GGA results are also listed, for a comparison, 
for the $2p$-states of the oxygen and nitrogen atoms and the TM atomic orbitals.

The pSIC method usually tends to localize the electrons\cite{psic-3} and
delocalize the holes.\cite{hole} With the DFT+U method,
the $3d$-TM orbital occupations often increase and the $sp$-orbital occupations of
the TM-neighbours decrease, with respect to the DFT result. 
For the pSIC method this is also usually true, due to the fact that
the self-interaction correction is stronger for the localized shells.
Netherveless, this is not a 'golden rule' and for some cases it might be broken.
The cases of porphyrin-V and -VO  are special.
Without oxygen, the $sp$-orbitals of V are substantially occupied,
because they overlap much with the neighbouring nitrogens.
It is interesting that with the pSIC aproach, 
the polarization of $sp$-V is antiferromagnetic (AF) to $3d$-V.
Large spin-polarizarions of nitrogens - around -0.3 $\mu_B$, which gives -1.2 $\mu_B$
in total - are not surprising.
Nitrogen is an element which usually couples antiferromagnetically to TM -
as for instance in (GaN,Mn) dilute magnetic semiconductor.\cite{Dublin} 
Moreover,
the $sp$-V orbitals overlap and couple ferromagnetically (FM) with the $sp$-N shells. 
The fact that the $4p$-shell of V in porphyrins is substantially occupied
can be compared to the similar effect observed in (GaAs,Mn)\cite{hole}
and Si:Mn.\cite{Stroppa}
On the other hand, in the GGA approach, the $sp$-V shell
couples ferromagnetically to $3d$-V, since the strong SI error is not
corrected for the $sp$-N orbitals.

For the porphyrin-VO, the $sp$-N is also AF-coupled with respect to $3d$-V.  
With the difference that the strong coupling of $sp$-V with N-$sp$ 
is replaced by the V interaction with oxygen, 
which is more strongly polarizable than nitrogen.
The spin polarization of oxygen is
around -1.6 $\mu_B$ and antiferromagnetic with respect to $3d$-V. 
The $sp$-N localized spin is -0.27 $\mu_B$, which is similar to that for porphyrin-V.
However, the largest spin polarization
of the $sp$-N orbitals is in the case of porphyrin-Mn, around -0.47 $\mu_B$.
Interestingly, the coupling of $sp$-N with respect to $3d$-TM is antiferromagnetic
not for all studied 2D networks - the exceptions are: 
porphyrin-Co, -Cu, -FeO, -CoO, -NiO and -CuO.

This is a consequence of the 4-fold crystal symmetry and the electronic filling of 
the $3d$-TM orbitals. 
The TM atoms possess valence electrons of the $s-$ and $d-$type 
and need to make the bonds with 4 N atoms and oxygen. Formation of the local magnetic moment
at the TM atom means less electrons involved in the chemical bonds, and a hole might be
delocalized over these bonds. The lack of TM electrons with the spin up at the TM-N bonds 
- for the cases with the d-shell less occupied than it is for Mn - causes the domination
of the spin down polarization at these bonds,
and the magnetic coupling of N atoms to the TM is antiferromagnetic.
The coupling changes signum with the growing number of electrons at the $d-$-shell of TM,
and the cases with Ni and Cu are ferromagnetically oriented to the N atoms. 
Large spin polarization of the $sp$-O orbitals is very promissing 
for tailoring spintronic devices.
It is worth to notice that the orientation of the spin coupling at oxygen 
with respect to $3d$-TM is usually the same as that of 
the coupling between $sp$-N and $3d$-TM.

Let us have a look at carbons of the porphyrin molecules and bridges; 
these adjacent to nitrogens, denoted C1,
and farther, denoted: C2, C3 and C4 - see the caption of Table~1 and Fig.~4.
The carbon atom denoted C1 bears a spin which couples AF to $3d$-V
(with local C1 moment of -0.12 $\mu_B$) for the VO case, 
and FM (0.21 $\mu_B$) for the porphyrin-V. Other spin polarizations at C1 also
deserve our attention: for porphyrin-Fe (0.17 $\mu_B$),  
P-Mn (0.09 $\mu_B$), P-FeO and P-Co (0.11 $\mu_B$), and P-CoO (0.12 $\mu_B$).
The carbon atoms denoted C2 polarize mostly in the case of porphyrin-Co
(0.08 $\mu_B$). 
Even far from the TM atom, C3 atoms, are spin-polarized around 0.04-0.07 $\mu_B$ for 
porphyrin-Mn, -Fe, -FeO, -Co, -CoO.
Remarkably, in porphyrin-V, the carbon atom of the acetylenic-bridge, namely C4, couples
antiferromagnetically to C1, and it is spin-polarized of about -0.1 $\mu_B$.

\begin{figure*}[ht]
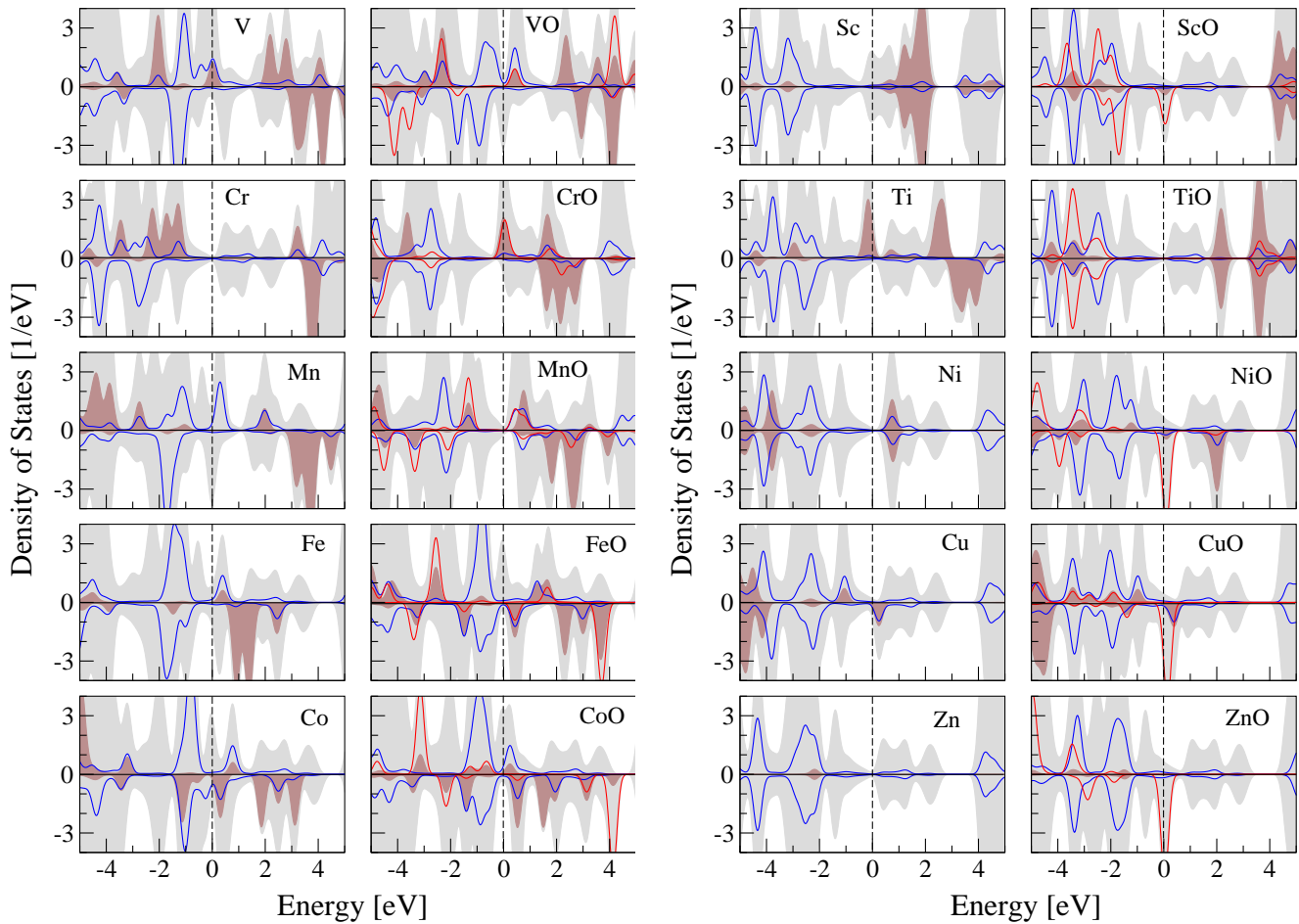

\centerline{
\includegraphics[scale=0.45,angle=0.0]{dos-psic-1.eps}\hspace{3mm}
\includegraphics[scale=0.45,angle=0.0]{dos-psic-2.eps}}
\caption{Projected densities of states (PDOS) of the
metal- and oxo-metal-porphyrin 2D networks
obtained with the pSIC approach. Grey color depicts the total DOS, the brown color
is the PDOS of $3d$-TM states, the blue line is the PDOS of $2p$-N (summed for 4 atoms),
the red line is the PDOS of $2p$-O states. Fermi level is at zero energy.}
\label{dpsic}
\end{figure*}

The results presented in this subsection align with 
the experimental and theoretical conclusions
of the work on porphyrin-Cr adsorbed at the Co surface.\cite{porf-Cr-Co}
Namely, the local spin-moments induced at nitrogens are AF-coupled to Cr, and 
the spin of less than halfly-occupied $3d$-shell of Cr is AF-coupled
to the more than halfly-occupied $3d$-shell of Co substrate - 
as in our case, it holds for the Cr and O localized magnetic moments.    

As for the difference between the pSIC and the GGA results, 
the pSIC magnetizations are always larger than the GGA ones. 
This is due to the fact that the Hund's rule is strongly obeyed when the electron 
localization is higher – like in the pSIC and DFT+U case.
Moreover, for the separate molecules, the magnetizations are localized only at 
the transition metals and oxygen. In contrast, for the periodic systems, the magnetizations 
are also partially localized on N, O, and even carbons of the acethylene bridges. 

Even if the magnetization is not localized at the acetylene bridges, the
covalent bonds between the molecules are essential for the spin delocalization
within the network-building molecules. 
Without these bonds, the molecules would be spin-polarized
with the well localized integer spin multiplicity, which would not be present
at the nitrogen atoms.
In contrast, for the periodic systems containing the N atoms and magnetic
impurities - such as already mentioned diluted magnetic semiconductor (Ga,Mn)N -   
the spin polarization is delocalized to the nitrogen atoms, with quite substantial
contribution.\cite{Dublin}

The more extensive view on the magnetization can be derived from 
the projected densities of states (PDOS) onto the $2p$-N, $3d$-TM
and $2p$-O states, and the total DOS, which are presented in Fig.~5; 
for the results obtained with the pSIC approach.
Similar PDOS plots for the GGA method are included in Fig.~S3 in the supplemental
information.\cite{supp} The numbers presented in Table~1 are obtained from the
quadrature of the atomic spin-densities presented in Fig.~5. It is in the energy range
from the bottom of the valence band - not presented in the figure - up to the Fermi
level. The spin-asymmetry of the PDOS is visible for the cases:  
V, Cr, Mn, Fe, and Co without and with the oxygen. 
The spin-asymmetry of the $2p$-N PDOS is pronounced even
for the cases where the L\"owdin spin-polarizations are rather small.
Interestingly, for some cases - like Cr - the spin polarization at the Fermi
level does not show up, but the total contribution to magnetism originates
from the deeper states. In this case, mostly from Cr and less from the nitrogen
atoms.

\subsection{Metallicity}

In Fig.~5, we see that most of P-TM and P-TMO networks are metallic. 
The Fermi levels of some
cases - namely Cr, MnO, TiO, Ni, Zn - are placed within a little energy gap.
The half-metallicity is plausible in the cases: Ti, Cu, VO, CrO, FeO, CoO, where
the more accurate GW-calculations would be necessary.\cite{GW}
It is very usual for the TM-doped metallic systems, that the Fermi level cuts the
$3d$-states. The effect of such methods like the DFT+U or the pSIC with respect to
the GGA results moves the $3d$-states
away from the Fermi level, if these states are not pinned to it.             
In our systems, however, some cases calculated with the GGA approach show clearly
the TM-based metallicity, while in the pSIC approach, the $3d$ states disappear
from the Fermi level - 
e.g. for the porphyrin-Mn, -Fe, -VO, -MnO, -CoO and -Ni networks.
Purely carbon-based metallicities in such magnetic systems 
like P-VO, P-Mn and P-FeO networks seem to be
promissing for spintronic devices, because the magnetic couplings in these cases
might be long range. Interestingly, there are also the metallic systems 
where the oxygen states dominate at the Fermi level. These systems are almost half metallic
- for instance in the porphyrin-NiO, -CuO and -ZnO networks.

\section{Conclusions}

2D organic networks are promissing for spintronic applications. 
In this work, we searched for magnets among 
oxo-metal-porphyrins and metal-porphyrins connected with the acetylene bridges.
For magnetism, the relative energetic positions of the $3d-$, $4s-$ and $4p-$ transition-metal
states with respect to the $2p$-states of the neighbouring atoms are very important.
Therefore, we used the self-interaction corrected pseudopotential scheme (pSIC), 
within the DFT framework for the calculations of the electronic structure.

Most of the porphyrin networks with TM and TMO are magnetic; except the Sc, TiO,
Ni and Zn cases. Addition of oxygen increases the magnetizations 
for the second-half $3d$-row TMs and the ScO case.
The rest of cases: Ti, V, Cr and Mn decrease magnetizations after the oxygen addition. 

Difference between the total and absolute magnetizations indicate ferrimagnetism,
which is the strongest for the porphyrin-V, -VO and -Mn networks. Further examination
of the systems with the L\"owdin populations tool shows the AF-coupling
of the oxygen magnetic moment with the TM local moment for the P-VO and P-CrO cases, 
and FM-coupling for the P-FeO, P-CoO, P-NiO, P-CuO and P-ZnO systems.
Interestingly, in the P-ScO case, the magnetic moment is localized at oxygen and not
at TM, and in the MnO case the situation is opposite. The strongest spin-polarization
localized at nitrogens, and AF-coupled to the $3d$-TM local moment, 
was found for the P-Mn, P-V, P-VO and P-Fe cases. The same couplings are ferromagnetic
for the P-NiO, P-Cu and P-CuO 2D networks. The spin-polarization effects are very long
range for the porphyrin networks with the TMs from V to Co. In these cases, 
the spin-asymmetry of the atomic shells was found very far from the TM atoms,
and it was well pronounced even on the acetylenic bridges.

In summary, the ferrimagnetic cases found in our studies are characterized by the
induced long-range spin-polarizations. 
These systems certainly would be good candidates for the high-T$_c$ 
magnets and will show up
new properties when are deposited at the surfaces of other materials. \\

{\bf Acknowledgements} \\

This work has been supported by the 
the National Science Center in Poland 
(the Project No. 2013/11/B/ST3/04041).
Calculations have been performed in the Interdisciplinary Centre of
Mathematical and Computer Modeling (ICM) of the University of Warsaw
within the grant G59-16. \\


\begin{thebibliography}{99}
\bibitem{supp}
[URL will be inserted by AIP] for [the details of the geometry and magnetizations
obtained within the GGA approach.]
\end{thebibliography}
\end{document}